# MODELING SEISMIC WAVE PROPAGATION IN TTI MEDIA USING RESIDUAL PERFECTLY MATCHED LAYER

Yuqin Luo, College of Science, Northeast Electric Power University, 169 Changchun Road, Chuanying District, Jilin 132012, luoyq1994@163.com.

Xintong Dong, College of Instrumentation and Electrical Engineering, Jilin University, 938 S, West Minzhu Ave, Changchun,130026, 18186829038@163.com.

Shiqi Dong, Department of Communication Engineering, College of Electric Engineering, Northeast Electric Power University, 169 Changchun Road, Chuanying District, Jilin, 132012, dsq1994@126.com.

Tie Zhong, Department of Communication Engineering, College of Electric Engineering, Northeast Electric Power University, 169 Changchun Road, Chuanying District, Jilin, 132012, 519647817@qq.com.

Yu Zhang, College of Science, Northeast Electric Power University, 169 Changchun Road, Chuanying District, Jilin, 132012,521zhangyu2008@163.com.

Ying Wang, the GBA Branch of Aerospace Information Research Institute, Chinese Academy of Sciences, 11 Kaiyuan Avenue, Huangpu District, Guangzhou 510530,wangying01@aircas.ac.cn.

Ning Hu, College of Mechanical Engineering, Hangzhou Dianzi University, 115 Wenyi Road, Xihu District, Hangzhou 310018, hooning@hdu.edu.cn.

# ABSTRACT

The perfectly matched layer(PML) is commonly used in wave propagation, radiation and diffraction problems in unbounded space domains. A new implementation scheme of PML is presented. The PML formulation is pre-defined, and the wave field absorption is achieved by calculating the residual between the PML equation and original equation through backward induction. Two forms of the Residual PML (RPML) are presented: RPML-I, which defines the residual as the difference between the original and PML equations, and RPML-II, which defines the residual as the difference between the original and PML wave fields. RPML-II is the simplest and easiest to extend, as it does not alter the original equation and only has one time partial derivative term in the residual equation. Additionally, since the residual equation has no spatial partial derivative term, high-order spatial difference discretization is unnecessary, which results in higher accuracy and computational efficiency. Furthermore, simulating a wave field in TTI media requires a high absorption effect and stability of PML. The numerical simulation demonstrates that RPML-II provides better absorption performance and stability compared to ADEPML and NPML. To meet the needs of wave field simulation for complex media, a multiaxial complex frequency shifted RPML-II (MCFS-RPML-II) is introduced, which employs double damping profiles and complex frequency shift technology to achieve higher stability and absorption effects.

# INTRODUCTION

How to construct effective boundary conditions to absorb the outgoing waves is one of the important challenges in unbounded domain wave field simulation. Several techniques have been developed in the last four decades: paraxial conditions(Clayton and Engquist, 1977; Enguist and Majda, 1977; Higdon, 1991), sponge zones (Cerjan et al.,1985; Sochacki et al., 1987), artificial transparent(Reynolds, 1978). The PML was proposed by Berenger and has been widely used in various fields, including, for example, the Helmholtz equation(Teixeira and Chew, 1998; Harari et al., 2000, Wu et al.,2022); linearized Euler equation(Hu, 1996; Nataf, 2005); Maxwell's equations(Berenger, 1994, 1996; Katz et al., 1994; Lei et al., 2022); Schrodinger equation(Nissen and Kreiss, 2011; Wang et al., 2006); nonlinear Klein-Gordon equations(Antoine and Zhao, 2022); High Power microwave (HPM) simulation(Zhang and Li, 2023), etc. It has also been applied successfully in seismic wave field simulation for acoustic(Lei et al, 2023), viscoelastic(Wang et al., 2019), elastic(Chew and Liu, 1996; Hastings et al., 1996), anisotropic(Collino and Tsogka, 2001) and porous media(Zeng et al., 2001), as well as full waveform inversion(Dong, 2020; Guo et al., 2024) and migration imaging(Wu and Dong, 2023; Dong, et al., 2024). A critical research focus for PML is how to avoid expensive temporal convolution operations. Among the numerous methods avoiding convolution operations, such as the classical variable splitting method (S-PML)(Hastings et al., 1996); the recursive convolution update approach (C-PML)(Appelö and Kreiss, 2006; Roden and Gedney, 2000; Komatitsch and Martin, 2007); the recursive time integration technology (RI-PML)(Drossaert and Giannopoulos, 2007; Giannopoulos, 2018); introducing auxiliary differential equations (ADE-PML)(Duru, 2014, 2015); the Z-transform and digital filtering methods(MZT-PML)(Ramadan and Oztoprak, 2002; Feng et al., 2015); the approximate transformation method (NPML)(Cummer, 2003; Luo and Liu, 2020, 2022). Compared to classical SPML, non-split PMLs have been developed that improve computational efficiency and save storage space.

PML performance is evaluated based on its absorption effect and stability. However, the PML can become unstable when simulating wave field propagation in certain specific situations, especially in anisotropic or dispersive media, e.g. in (visco)elastodynamics(Appelö and Kreiss, 2006; Komatitsch and Martin, 2007; Hu, 2001; Bécache et al., 2003; Li et al., 2019; Appelö and Colonius; 2009), computational electrom-agnetics(Bécache et al., 2017a, 2017b; Cummer, 2004), and aeroacoustics(Hu, 2001; Demaldent and Imperiale, 2013). To address this issue, Meza-Fajardo proved that traditional PML fails to satisfy strict asymptotic stability in both isotropic and anisotropic media and proposed the multi-axis perfect Matching Layer(M-SPML) based on SPML(Mezafajardo and Papageorgiou, 2008, 2010). Additionally, when propagation occurs at a near-grazing incident angle or in a shallow layer, the PML attenuation factor based on CCS transformation may yield small integral values, making it challenging to absorb wave field energy effectively. Furthermore, the CCS transform formula has a frequency variable in the denominator, leading to singular values at extremely low frequencies. Therefore, Kuzuoglu modified the form of complex coordinate stretching transformation and proposed the complex frequency shifted(CFS) technology(Kuzuoglu and Mittra, 1996; Ma et al., 2018, 2019; Huang et al., 2023). The CFS formula introduces frequency shift factor and scale factor, which play the role of eliminating singular values and bending the wave field to the normal direction, respectively. Although CFS transform was not directly applied to SPML due to technical limitations at the time, its introduction has paved the way for more advanced PML technologies.

Different convolution avoidance strategies affect the computational efficiency of PML, Multiaxial strategy is an effective means to solve the stability problem; complex frequency shift transformation is the primary choice to enhance the absorption of grazing and evanescent waves, and can further improve the stability to a certain extent. This paper provides a new PML(residual PML), which is built by calculating the residual in reverse. RPML is a non-splitting method, with the simplest auxiliary equation form and only one time derivative term, which has higher computational efficiency. Compared with NPML and ADEPML, the residual perfectly matched layer has better absorption effect. Based on RPML, this article further realizes complex frequency shift transformation and multi-axis attenuation processing, and the multiaxial complex frequency shifted residual perfectly matched layer(MCFS-RPML) has higher stability and absorption performance.

## METHODS

**Governing equations**

In this paper, the wave field simulation is carried out using the time second-order space-12$^{th}$-order staggered grid finite difference algorithm. The first-order velocity stress equations are as follows:

$$\begin{aligned} \rho \partial_t \boldsymbol{v} &= \nabla \cdot \boldsymbol{\tau} \\ \partial_t \boldsymbol{\tau} &= \mathbf{C} : \nabla \boldsymbol{v} \end{aligned} \quad (1)$$

Where C is the elastic coefficient tensor matrix, v is the velocity component, τ is the stress tensor, ρ is the mass density. The $C_{TTI}$ can be obtained by rotating the elastic matrix of the VTI through a bond transformation by a certain angle:

$$C_{VTI} = \begin{bmatrix} C_{11} & C_{12} & C_{13} & 0 & 0 & 0 \\ C_{21} & C_{22} & C_{23} & 0 & 0 & 0 \\ C_{31} & C_{32} & C_{33} & 0 & 0 & 0 \\ 0 & 0 & 0 & C_{44} & 0 & 0 \\ 0 & 0 & 0 & 0 & C_{55} & 0 \\ 0 & 0 & 0 & 0 & 0 & C_{66} \end{bmatrix} \quad (2)$$

$$C_{TTI} = M_\phi M_\theta C_{TTI} M_\theta^T M_\phi^T = \begin{bmatrix} C_{11} & C_{12} & C_{13} & C_{14} & C_{15} & C_{16} \\ C_{21} & C_{22} & C_{23} & C_{24} & C_{25} & C_{26} \\ C_{31} & C_{32} & C_{33} & C_{34} & C_{35} & C_{36} \\ C_{41} & C_{42} & C_{43} & C_{44} & C_{45} & C_{46} \\ C_{51} & C_{52} & C_{53} & C_{54} & C_{55} & C_{56} \\ C_{61} & C_{62} & C_{64} & C_{64} & C_{65} & C_{66} \end{bmatrix} \quad (3)$$

$\theta$ and $\varphi$ are the polarization and azimuth angles, respectively, and $M_\theta$ and $M_\varphi$ are defined as follows:

$$M_\theta = \begin{bmatrix} \cos^2\theta & 0 & \sin^2\theta & 0 & -\sin 2\theta & 0 \\ 0 & 1 & 0 & 0 & 0 & 0 \\ \sin^2\theta & 0 & \cos^2\theta & 0 & \sin 2\theta & 0 \\ 0 & 0 & 0 & \cos\theta & 0 & \sin\theta \\ \frac{1}{2}\sin 2\theta & 0 & -\frac{1}{2}\sin 2\theta & 0 & \cos 2\theta & 0 \\ 0 & 0 & 0 & -\sin\theta & 0 & \cos\theta \end{bmatrix} M_\phi = \begin{bmatrix} \cos^2\phi & \sin^2\phi & 0 & 0 & 0 & -\sin 2\phi \\ \sin^2\phi & \cos^2\phi & 0 & 0 & 0 & \sin 2\phi \\ 0 & 0 & 1 & 0 & 0 & 0 \\ 0 & 0 & 0 & \cos\phi & \sin\phi & 0 \\ 0 & 0 & 0 & -\sin\phi & \cos\phi & 0 \\ \frac{1}{2}\sin 2\phi & -\frac{1}{2}\sin 2\phi & 0 & 0 & 0 & \cos 2\phi \end{bmatrix} \quad (4)$$

The system of velocity stress equations is converted to the frequency domain, subjected to complex coordinate stretching transformation, and then converted back to time domain to obtain the new governing equations after loading the PML. Translating Eqs. 1 to frequency domain:

$$\begin{aligned} i\omega\rho\tilde{v} &= \nabla \cdot \tilde{\tau} \\ i\omega\tilde{\tau} &= \mathbf{C} : \nabla\tilde{v} \end{aligned} \quad (5)$$

Where $\tilde{v}$ is the velocity component in frequency domain, $\tilde{\tau}$ is the stress tensor in frequency domain. The complex coordinate stretching formula is

$$\begin{aligned} \tilde{\partial z} &= \frac{1}{S_z}\partial z, S_z = 1 + \frac{\alpha_z}{i\omega} \\ \tilde{\partial x} &= \frac{1}{S_x}\partial x, S_x = 1 + \frac{\alpha_x}{i\omega} \end{aligned} \quad (6)$$

Substituting Eqs. 6 into 5,

$$\begin{aligned}
i\omega\rho\tilde{v}_x &= S_x^{-1}\partial_x\tilde{\tau}_{xx} + S_z^{-1}\partial_z\tilde{\tau}_{zx} \\
i\omega\rho\tilde{v}_z &= S_x^{-1}\partial_x\tilde{\tau}_{zx} + S_z^{-1}\partial_z\tilde{\tau}_{zz} \\
i\omega\tilde{\tau}_{xx} &= C_{11}S_x^{-1}\partial_x\tilde{v}_x + C_{13}S_z^{-1}\partial_z\tilde{v}_z + C_{15}\left(S_z^{-1}\partial_z\tilde{v}_x + S_x^{-1}\partial_x\tilde{v}_z\right) \\
i\omega\tilde{\tau}_{zz} &= C_{13}S_x^{-1}\partial_x\tilde{v}_x + C_{33}S_z^{-1}\partial_z\tilde{v}_z + C_{35}\left(S_z^{-1}\partial_z\tilde{v}_x + S_x^{-1}\partial_x\tilde{v}_z\right) \\
i\omega\tilde{\tau}_{zx} &= C_{15}S_x^{-1}\partial_x\tilde{v}_x + C_{35}S_z^{-1}\partial_z\tilde{v}_z + C_{55}\left(S_z^{-1}\partial_z\tilde{v}_x + S_x^{-1}\partial_x\tilde{v}_z\right)
\end{aligned} \qquad (7)$$

In the process of translating Eqs. 7 to time domain, how to compute convolution terms or avoid convolution operations is the focus of PML research, we can divide them into two categories: split method (SPML) and non split methods(CPML, ADEPML, RIPML and NPML).

After transformation, the system of new equations or wave field variables can be expressed in terms of the original plus a residual term. Therefore, The residual ε can be defined as the difference between the original equations and the PML equations (Eqs. 8) or the difference between the original wave field and PML wave field (Eqs. 9) :

$$\begin{aligned}
i\omega\rho\tilde{\mathbf{v}} &= \nabla\cdot\tilde{\boldsymbol{\tau}} - \tilde{\boldsymbol{\varepsilon}} \\
i\omega\tilde{\boldsymbol{\tau}} &= \mathbf{C}:\left(\nabla\tilde{\mathbf{v}} - \tilde{\boldsymbol{\varepsilon}}\right)
\end{aligned} \qquad (8)$$

and

$$\begin{aligned}
i\omega\rho\tilde{\mathbf{v}} &= \nabla\cdot\left(\tilde{\boldsymbol{\tau}} - \tilde{\boldsymbol{\varepsilon}}\right) \\
i\omega\tilde{\boldsymbol{\tau}} &= \mathbf{C}:\nabla\left(\tilde{\mathbf{v}} - \tilde{\boldsymbol{\varepsilon}}\right)
\end{aligned} \qquad (9)$$

Converting back to the time domain,

$$\begin{aligned}
\partial_t\rho\mathbf{v} &= \nabla\cdot\boldsymbol{\tau} - \boldsymbol{\varepsilon} \\
\partial_t\boldsymbol{\tau} &= \mathbf{C}:\left(\nabla\mathbf{v} - \boldsymbol{\varepsilon}\right)
\end{aligned} \qquad (10)$$

and

$$\begin{aligned}
\partial_t\rho\mathbf{v} &= \nabla\cdot\left(\boldsymbol{\tau} - \boldsymbol{\varepsilon}\right) \\
\partial_t\boldsymbol{\tau} &= \mathbf{C}:\nabla\left(\mathbf{v} - \boldsymbol{\varepsilon}\right)
\end{aligned} \qquad (11)$$

The RPML-I and RPML-II equations (Eqs. 10 and 11 respectively) have distinct physical interpretations, giving rise to a unique classification system. The fundamental principle underpinning the RPML is to calculate the residual, which we can obtain by direct method and backward induction method. The backward induction method calculates the residual through the preexisting PML(ADE-PML /RIPML/NPML) in reverse, and we can take this as an opportunity to study the connections between different PMLs.

**RPML-I residual equations**

Assuming Eqs. 8 are known, Eqs. 7 and 8 are equivalent, and we get

$$\begin{aligned}
&\partial_x \tilde{\tau}_{xx} + \partial_z \tilde{\tau}_{zx} - \tilde{\varepsilon}_{xx}^x - \tilde{\varepsilon}_{zx}^z = S_x^{-1} \partial_x \tilde{\tau}_{xx} + S_z^{-1} \partial_z \tilde{\tau}_{zx} \\
&\partial_x \tilde{\tau}_{zx} + \partial_z \tilde{\tau}_{zz} - \tilde{\varepsilon}_{zx}^x - \tilde{\varepsilon}_{zz}^z = S_x^{-1} \partial_x \tilde{\tau}_{zx} + S_z^{-1} \partial_z \tilde{\tau}_{zz} \\
&C_{11}\left(\partial_x \tilde{v}_x - \tilde{\varepsilon}_x^x\right) + C_{13}\left(\partial_z \tilde{v}_z - \tilde{\varepsilon}_z^z\right) + C_{15}\left(\partial_z \tilde{v}_x + \partial_x \tilde{v}_z - \tilde{\varepsilon}_z^x - \tilde{\varepsilon}_x^z\right) \\
&\quad = C_{11} S_x^{-1} \partial_x \tilde{v}_x + C_{13} S_z^{-1} \partial_z \tilde{v}_z + C_{15}\left(S_z^{-1} \partial_z \tilde{v}_x + S_x^{-1} \partial_x \tilde{v}_z\right) \\
&C_{13}\left(\partial_x \tilde{v}_x - \tilde{\varepsilon}_x^x\right) + C_{33}\left(\partial_z \tilde{v}_z - \tilde{\varepsilon}_z^z\right) + C_{35}\left(\partial_z \tilde{v}_x + \partial_x \tilde{v}_z - \tilde{\varepsilon}_z^x - \tilde{\varepsilon}_x^z\right) \\
&\quad = C_{13} S_x^{-1} \partial_x \tilde{v}_x + C_{33} S_z^{-1} \partial_z \tilde{v}_z + C_{35}\left(S_z^{-1} \partial_z \tilde{v}_x + S_x^{-1} \partial_x \tilde{v}_z\right) \\
&C_{15}\left(\partial_x \tilde{v}_x - \tilde{\varepsilon}_x^x\right) + C_{35}\left(\partial_z \tilde{v}_z - \tilde{\varepsilon}_z^z\right) + C_{55}\left(\partial_z \tilde{v}_x + \partial_x \tilde{v}_z - \tilde{\varepsilon}_z^x - \tilde{\varepsilon}_x^z\right) \\
&\quad = C_{15} S_x^{-1} \partial_x \tilde{v}_x + C_{35} S_z^{-1} \partial_z \tilde{v}_z + C_{55}\left(S_z^{-1} \partial_z \tilde{v}_x + S_x^{-1} \partial_x \tilde{v}_z\right)
\end{aligned} \quad (12)$$

Taking stress component as an example,

$$\begin{aligned}
S_x\left(\partial_x \tilde{\tau}_{xx} - \tilde{\varepsilon}_{xx}^x\right) &= \partial_x \tilde{\tau}_{xx} \\
S_z\left(\partial_z \tilde{\tau}_{zx} - \tilde{\varepsilon}_{zx}^z\right) &= \partial_z \tilde{\tau}_{zx}
\end{aligned} \quad (13)$$

Substituting Eqs. 6 into 13, one obtains:

$$\begin{aligned}
\alpha_x \partial_x \tilde{\tau}_{xx} &= i\omega \tilde{\varepsilon}_{xx}^x + \alpha_x \tilde{\varepsilon}_{xx}^x; \\
\alpha_z \partial_z \tilde{\tau}_{zx} &= i\omega \tilde{\varepsilon}_{zx}^z + \alpha_z \tilde{\varepsilon}_{zx}^z
\end{aligned} \quad (14)$$

Converting back to the time domain, one obtains:

$$\begin{aligned}
&\alpha_m \partial_m \xi = \partial_t \varepsilon^m + \alpha_m \varepsilon^m \\
&\varepsilon^m : \varepsilon_{xx}^x, \varepsilon_{zx}^m, \varepsilon_{zz}^z, \varepsilon_x^m, \varepsilon_z^m \\
&\xi : \tau_{xx}, \tau_{zz}, \tau_{zx}, v_x, v_z \\
&m : x, z
\end{aligned} \quad (15)$$

One can notice that Eqs. 14 has the same form as the ADE-PML formulation. The ideas of the auxiliary memory variables developed for Maxwell's equations by Gedney and Zhao

(2010).

**RPML-II residual equations**

RPML-I is equivalent to ADEPML, while RPML-II is the focus of this study. We obtain residuals through direct and backward derivation methods. The backward derivation method is to obtain the residual in reverse through three commonly used PMLs.

*Direct method*

Combining Eqs. 7 and 9 yields:

$$\begin{aligned}
&\partial_x\left(\tilde{\tau}_{xx} - \tilde{\varepsilon}_{xx}^x\right) + \partial_z\left(\tilde{\tau}_{zx} - \tilde{\varepsilon}_{zx}^z\right) = S_x^{-1}\partial_x\tilde{\tau}_{xx} + S_z^{-1}\partial_z\tilde{\tau}_{zx} \\
&\partial_x\left(\tilde{\tau}_{zx} - \tilde{\varepsilon}_{zx}^x\right) + \partial_z\left(\tilde{\tau}_{zz} - \tilde{\varepsilon}_{zz}^z\right) = S_x^{-1}\partial_x\tilde{\tau}_{zx} + S_z^{-1}\partial_z\tilde{\tau}_{zz} \\
&C_{11}\partial_x\left(\tilde{v}_x - \tilde{\varepsilon}_x^x\right) + C_{13}\partial_z\left(\tilde{v}_z - \tilde{\varepsilon}_z^z\right) + C_{15}\left[\partial_z\left(\tilde{v}_x - \tilde{\varepsilon}_x^z\right) + \partial_x\left(\tilde{v}_z - \tilde{\varepsilon}_z^x\right)\right] \\
&= C_{11}S_x^{-1}\partial_x\tilde{v}_x + C_{13}S_z^{-1}\partial_z\tilde{v}_z + C_{15}\left(S_z^{-1}\partial_z\tilde{v}_x + S_x^{-1}\partial_x\tilde{v}_z\right) \\
&C_{13}\partial_x\left(\tilde{v}_x - \tilde{\varepsilon}_x^x\right) + C_{33}\partial_z\left(\tilde{v}_z - \tilde{\varepsilon}_z^z\right) + C_{35}\left[\partial_z\left(\tilde{v}_x - \tilde{\varepsilon}_x^z\right) + \partial_x\left(\tilde{v}_z - \tilde{\varepsilon}_z^x\right)\right] \\
&= C_{13}S_x^{-1}\partial_x\tilde{v}_x + C_{33}S_z^{-1}\partial_z\tilde{v}_z + C_{35}\left(S_z^{-1}\partial_z\tilde{v}_x + S_x^{-1}\partial_x\tilde{v}_z\right) \\
&C_{15}\partial_x\left(\tilde{v}_x - \tilde{\varepsilon}_x^x\right) + C_{35}\partial_z\left(\tilde{v}_z - \tilde{\varepsilon}_z^z\right) + C_{55}\left[\partial_z\left(\tilde{v}_x - \tilde{\varepsilon}_x^z\right) + \partial_x\left(\tilde{v}_z - \tilde{\varepsilon}_z^x\right)\right] \\
&= C_{15}S_x^{-1}\partial_x\tilde{v}_x + C_{35}S_z^{-1}\partial_z\tilde{v}_z + C_{55}\left(S_z^{-1}\partial_z\tilde{v}_x + S_x^{-1}\partial_x\tilde{v}_z\right)
\end{aligned} \quad (16)$$

Substituting complex coordinate stretching formula into Eqs. 16,

$$\begin{aligned}
S_x\partial_x\left(\tilde{\tau}_{xx} - \tilde{\varepsilon}_{xx}^x\right) &= \partial_x\tilde{\tau}_{xx} \\
S_z\partial_z\left(\tilde{\tau}_{zx} - \tilde{\varepsilon}_{zx}^z\right) &= \partial_z\tilde{\tau}_{zx}
\end{aligned} \quad (17)$$

$$i\omega\partial_x\tilde{\varepsilon}_{xx}^x = \alpha_x\partial_x\left(\tilde{\tau}_{xx} - \tilde{\varepsilon}_{xx}^x\right); i\omega\partial_z\tilde{\varepsilon}_{zx}^z = \alpha_z\partial_z\left(\tilde{\tau}_{zx} - \tilde{\varepsilon}_{zx}^z\right) \quad (18)$$

Eqs. 18 is more complicated than Eqs. 15, and we take Eqs. 18 a step further by approximation:

$$i\omega\partial_x\tilde{\varepsilon}_{xx}^x = \partial_x\left[\alpha_x\left(\tilde{\tau}_{xx} - \tilde{\varepsilon}_{xx}^x\right)\right]; i\omega\partial_z\tilde{\varepsilon}_{zx}^z = \partial_z\left[\alpha_z\left(\tilde{\tau}_{zx} - \tilde{\varepsilon}_{zx}^z\right)\right] \quad (19)$$

This approximate treatment was applied by Cummer to derive NPML(Cummer, 2003).

Finally, we can get the residual equation of RPML-II:

$$\partial_t \varepsilon^m = \alpha_m \left( \xi - \varepsilon^m \right) \quad (20)$$

Eq. 20 is the residual formula of RPML-II, which can also be obtained by inverse calculation of other PMLs.

*Backward induction method (RI-RPML)*

In this part, we take the absorbing wave field under RIPML as the final wave field of RPML-II. The basic principle of the RIPML is to rewrite the velocity-stress Eqs 6 by introducing two new auxiliary tensors, we denote the components as

$$\begin{aligned}
\tilde{M}_x^x &= S_x^{-1} \partial_x \tilde{v}_x ; \tilde{M}_z^z = S_z^{-1} \partial_z \tilde{v}_z \\
\tilde{M}_z^x &= S_x^{-1} \partial_x \tilde{v}_z ; \tilde{M}_x^z = S_z^{-1} \partial_z \tilde{v}_x \\
\tilde{M}_{zz}^z &= S_x^{-1} \partial_z \tilde{\tau}_{zz} ; \tilde{M}_{zx}^x = S_z^{-1} \partial_x \tilde{\tau}_{zx} \\
\tilde{M}_{xx}^x &= S_x^{-1} \partial_x \tilde{\tau}_{xx} ; \tilde{M}_{zx}^z = S_z^{-1} \partial_z \tilde{\tau}_{zx}
\end{aligned} \quad (21)$$

Substituting the above equations into Eqs. 7 and converting back to time domain,

$$\begin{aligned}
\rho \partial_t v_x &= M_{xx}^x + M_{zx}^z \\
\rho \partial_t \tilde{v}_z &= M_{zx}^x + M_{zz}^z \\
\partial_t \tilde{\tau}_{xx} &= C_{11} M_x^x + C_{13} M_z^z + C_{15} \left( M_x^z + M_z^x \right) \\
\partial_t \tilde{\tau}_{zz} &= C_{13} M_x^x + C_{33} M_z^z + C_{35} \left( M_x^z + M_z^x \right) \\
\partial_t \tilde{\tau}_{zx} &= C_{15} M_x^x + C_{35} M_z^z + C_{55} \left( M_x^z + M_z^x \right)
\end{aligned} \quad (22)$$

Substituting the complex stretch transformation function, Eqs. 21 is transformed into time domain in integrated form, abbreviated as

$$\begin{aligned}
\int_0^t \alpha_m M_n^m dt &= \partial_m v_n - M_n^m ; \\
\int_0^t \alpha_m M_{mn}^m dt &= \partial_m \tau_{mn} - M_{mn}^m \\
m,n &: x,z
\end{aligned} \quad (23)$$

Combining the RPML Eqs.11 with RIPML Eqs. 22, establish the relationship equation between the RIPML auxiliary variable and RPML residual, as follows:

$$M_m^n = \partial_n(v_m - \varepsilon_m^n); M_{mn}^n = \partial_n(\tau_{mn} - \varepsilon_{mn}^n) \tag{24}$$

Substituting Eqs. 24 into 23,

$$\int_0^t \alpha_m \partial_m (v_n - \varepsilon_n^m) dt = \partial_m \varepsilon_n^m$$
$$\int_0^t \alpha_m \partial_m (\tau_{mn} - \varepsilon_{mn}^m) dt = \partial_m \varepsilon_{mn}^m \tag{25}$$

If the approximate treatment used in Eqs. 19 is taken, Eqs. 25 can be rewritten as

$$\int_0^t \alpha_m (\xi - \varepsilon^m) dt = \varepsilon^m \tag{26}$$

Eq. 26 is a simplified residual equation obtained through approximate processing, which is equivalent to Eq. 20. Eq. 23 is the auxiliary equation of RIPML, which comprises an integral term and a partial derivative term in space. Utilizing the trapezoidal formula, we can handle the integral term, whereas the partial derivative term is discretized using the difference scheme. However, Eq. 26 has only one time integral term, and a more concise processing formula can bring higher computational efficiency compared with RIPML. We can adopt the trapezoidal approximation consistent with RIPML to solve Eq. 26 and obtain the recursive integral residual PML (RI-RPML).

*Backward induction method (ADE-RPML)*

Deriving the residual equation through the existing non-split PMLs is the core idea of the backward induction method, and ADEPML is one of the most widely used boundary conditions in seismic exploration.

We treat the complex stretching transform term in Eq. 7 as follows:

$$S_m^{-1}\partial_m\tilde{v}_n = \partial_m\tilde{v}_n - \frac{\alpha_m}{i\omega+\alpha_m}\partial_m\tilde{v}_n; S_m^{-1}\partial_m\tilde{\tau}_{nm} = \partial_m\tilde{\tau}_{nm} - \frac{\alpha_m}{i\omega+\alpha_m}\partial_m\tilde{\tau}_{nm} \quad (27)$$

Define auxiliary memory variable Q :

$$\tilde{Q}_m^{v_n} = -\frac{\alpha_m}{i\omega+\alpha_m}\partial_m\tilde{v}_n; \tilde{Q}_m^{\tau_{nm}} = -\frac{\alpha_m}{i\omega+\alpha_m}\partial_m\tilde{\tau}_{nm} \quad (28)$$

Substituting Q into Eqs. 7 and converting back to time domain, one obtains:

$$\begin{aligned}
\partial_t\rho v_x &= \partial_x\tau_{xx} + \partial_z\tau_{zx} + Q_x^{\tau_{xx}} + Q_z^{\tau_{zx}} \\
\partial_t\rho v_z &= \partial_x\tau_{zx} + \partial_z\tau_{zz} + Q_x^{\tau_{zx}} + Q_z^{\tau_{zz}} \\
\partial_t\tau_{xx} &= C_{11}\left(\partial_x v_x + Q_x^{v_x}\right) + C_{13}\left(\partial_z v_z + Q_z^{v_z}\right) + C_{15}\left(\partial_z v_x + \partial_x v_z + Q_x^{v_z} + Q_z^{v_x}\right) \\
\partial_t\tau_{zz} &= C_{13}\left(\partial_x v_x + Q_x^{v_x}\right) + C_{33}\left(\partial_z v_z + Q_z^{v_z}\right) + C_{35}\left(\partial_z v_x + \partial_x v_z + Q_x^{v_z} + Q_z^{v_x}\right) \\
\partial_t\tau_{zx} &= C_{15}\left(\partial_x v_x + Q_x^{v_x}\right) + C_{35}\left(\partial_z v_z + Q_z^{v_z}\right) + C_{55}\left(\partial_z v_x + \partial_x v_z + Q_x^{v_z} + Q_z^{v_x}\right)
\end{aligned} \quad (29)$$

Equation 29 is consistent with RPML- I, therefore, deriving the residual equation of RPMLII based on ADEPML is also equivalent to deriving that based on RPML - I. Combining Eqs.10, 27, and 29, the relationship between residual ε and auxiliary variable Q is as follows:

$$\begin{aligned}
\partial_m\tilde{\varepsilon}_n^m &= -\tilde{Q}_m^{v_n} = \frac{\alpha_m}{i\omega+\alpha_m}\partial_m\tilde{v}_n \\
\partial_m\tilde{\varepsilon}_{nm}^m &= -\tilde{Q}_m^{\tau_{nm}} = \frac{\alpha_m}{i\omega+\alpha_m}\partial_m\tilde{\tau}_{nm}
\end{aligned} \quad (30)$$

Converting back to time domain, one obtains:

$$\begin{aligned}
\partial_t\partial_m\varepsilon_n^m + \alpha_m\partial_m\varepsilon_n^m &= \alpha_m\partial_m v_n \\
\partial_t\partial_m\varepsilon_{nm}^m + \alpha_m\partial_m\varepsilon_{nm}^m &= \alpha_m\partial_m\tau_{nm}
\end{aligned} \quad (31)$$

Eqs. 31 is the residual equation obtained based on ADEPML (ADE-RPML), which has mixed partial derivatives of time and space and is difficult to deal with. However, by observing the characteristics of the equation, further approximation can be used to obtain a more concise residual equation:

$$\begin{aligned}
\partial_t\varepsilon_n^m &= \alpha_m\left(v_n - \varepsilon_n^m\right) \\
\partial_t\varepsilon_{nm}^m &= \alpha_m\left(\tau_{nm} - \varepsilon_{nm}^m\right)
\end{aligned} \quad (32)$$

The same result can be obtained by taking the time partial derivative of Eqs. 25 or time integration of both sides of Eqs. 31. Therefore, in theory, the residual of the two is the same, and the difference in actual absorption effect is only related to the discrete method.

*Backward induction method （N-RPML）*

NPML directly transforms the wave field, which has advantages in computation and promotion applications. NPML takes an approximate treatment of Eq.7 by placing the complex stretching term in the partial derivative term.

$$\begin{aligned}
i\omega\rho\tilde{v}_x &= \partial_x S_x^{-1}\tilde{\tau}_{xx} + \partial_z S_z^{-1}\tilde{\tau}_{zx} \\
i\omega\rho\tilde{v}_z &= \partial_x S_x^{-1}\tilde{\tau}_{zx} + \partial_z S_z^{-1}\tilde{\tau}_{zz} \\
i\omega\tilde{\tau}_{xx} &= C_{11}\partial_x S_x^{-1}\tilde{v}_x + C_{13}\partial_z S_z^{-1}\tilde{v}_z + C_{15}\left(\partial_z S_z^{-1}\tilde{v}_x + \partial_x S_x^{-1}\tilde{v}_z\right) \\
i\omega\tilde{\tau}_{zz} &= C_{13}\partial_x S_x^{-1}\tilde{v}_x + C_{33}\partial_z S_z^{-1}\tilde{v}_z + C_{35}\left(\partial_z S_z^{-1}\tilde{v}_x + \partial_x S_x^{-1}\tilde{v}_z\right) \\
i\omega\tilde{\tau}_{zx} &= C_{15}\partial_x S_x^{-1}\tilde{v}_x + C_{35}\partial_z S_z^{-1}\tilde{v}_z + C_{55}\left(\partial_z S_z^{-1}\tilde{v}_x + \partial_x S_x^{-1}\tilde{v}_z\right)
\end{aligned} \quad (33)$$

Define new variable as follows:

$$S_x^{-1}\tilde{\xi} = \bar{\tilde{\xi}} \quad (34)$$

Replacing the complex coordinate stretching formula and converting back to time domain, one obtains:

$$\partial_t \bar{\xi} + \alpha_m \bar{\xi} = \partial_t \xi \quad (35)$$

Eq. 35 is the NPML auxiliary variable equation. The new variable is calculated by Eq. 35, and then substituted into the original equation to realize the wave absorption. Combining Eqs. (9,33) yields:

$$\partial_t \varepsilon^m = \alpha_m \left(\xi - \varepsilon^m\right) \quad (36)$$

Eq. 36 is the residual equation obtained based on NPML, which is consistent with Eq. 20.

For NPML, the time partial derivatives of the new and original variable need to be discretized simultaneously, which makes it challenging to extend to higher order time.

Conversely, RPML-II offers the most concise form of implementation for discrete iteration, with only one time partial derivative term. Compared to auxiliary equations of other PMLs, the RPML-II-based auxiliary equation is the most concise and intuitively shows that it has the optimal computing performance.

RPML-II, like NPML, does not alter the governing equation but introduces a residual equation to obtain a new variable. After replacing the original variable, the wave field absorbed at the boundary can be obtained, which is easier to generalize to more complex medium wave field simulations. This advantage can be known from the derivation results. In the subsequent process of introducing CFS transformation, the auxiliary equations of all PMLs become extremely complex because of the introduction of frequency shift factor and scale factor, and even the introduction is difficult for some PMLs, such as SPML. The CFS technology is essential for complex wave field absorption. After introducing, the new variable equation becomes more complex in NPML, and the same is true for RPML-II. However, we obtain a new CFS-RPML-II residual equation through scale factor extraction, which is also the most concise equation among all kinds of CFS-PMLs. The auxiliary equation form is still consistent with Eq. 36, so both CCS-RPML-II and CFS-RPML-II are the most concise PML and the easiest to generalize. The computational efficiency and absorption effect are also verified in the subsequent actual seismic wave field simulation.

In this section, we successively establish the connection between ADEPML, RIPML, NPML, RPML I, and RPML II in the process of deriving RPML I and RPML II, as shown in Figure 1.

Figure 1 is the roadmap of the PML loading strategy. After CCS or CFS transformation in the frequency domain, the wave field is dissipated in PML. In general, the PML is based on CCS transformation. If it is based on CFS transformation, we call it CFS-PML. Afterwards, the equation is transformed into the time domain through inverse Fourier transform, during which various PMLs are derived based on different convolution processing methods. Different convolution processing methods will have a significant impact on the computational efficiency, stability and absorption effect, which is the core of the PML research.

Figure 1 shows the relationship between various PMLs and RPML involved in this study. CPML has been proven to be a special case of ADE-PML, The CPML can be seen as a particular case of the ADE-PML at the second order in time(Martin and Komatitsch, 2010), and ADEPML shares the same equation form as RPML - I. If the residual equation based on ADE-PML is integrated, the equation based on RIPML can be obtained. In theory, the residuals of the two are the same, indicating that their performance is actually similar. However, in practice, the differences in absorption effect, computational efficiency, and stability are due to their different discretization methods.

Of course, further approximation is applied to the residual equation obtained based on ADEPML and RIPML, and the resulting equation is consistent with the residual equation obtained based on NPML. This is also the difference between NPML and other PMLs. However, the approximate processing does not cause stronger false reflection, and has better absorption performance compared with CPML (Chen, 2012).

In the process of inverse extrapolation, we use the discrete methods of

ADEPML, RIPML and NPML to obtain different residual calculation schemes, which are called ADE-RPML-II, RI-RPML-II and N-RPML-II. In this paper, N-RPML-II is mainly used to simulate the wave field, which is consistent with the RPML-II derived directly.

**Multiaxial complex frequency shifted-RPML-II**

The CFS transformation introduces the frequency shift factor and scale factor to improve the absorption effect and stability. However, CFS-PML still has instability in wave simulation of anisotropic media, so we need to further introduce the multi-axis technology, and propose MCFS-RPML-II.

$$CCS: \partial \bar{m} = S_m^{-1} \partial m, S_m = 1 + \frac{\alpha_m}{i\omega}$$
$$CFS: \partial \bar{m} = S_m^{-1} \partial m, S_m = \beta_m + \frac{\alpha_m}{i\omega + \eta_m}$$
(37)

Where $\beta_m$ is the scale factor, $\eta_m$ is the frequency shift factor, and $\alpha_m$ is the attenuation factor.

The CFS-PML can effectively absorb the grazing wave and transient wave, and the stability is also enhanced because the dissipation wave in the shallow boundary layer is effectively absorbed.

As shown in Figure 2, the MCFS-RPML-II due to the introduction of the stability factor, the attenuation function is no longer equal to zero along the direction parallel to boundary. The double attenuation profiles can better absorb the wave field inside the boundary, so that the PML satisfies the asymptotic stability. However, the combination of CFS and multi-axis technology requires redefining scale factors and frequency shift factors at the boundaries. In region 1, the scale factor and frequency shift factor are consistent with CFS-RPML-II in the main direction, but in the parallel direction, the scale factor is set to 1 and the frequency shift

factor is set to 0, which degenerates into CCS transformation. The formula is as follows:

$$\text{domain 1}: S_m = 1 + \frac{\alpha_m}{i\omega} \qquad (38)$$
$$\text{domain 2}: S_x = \beta_x + \frac{\alpha_x^x}{i\omega + \eta_x}, S_z = 1 + \frac{P^{z/x}\alpha_x^x}{i\omega}$$

Domain 1 displays conventional RPML, while domain 2 displays MCFS-RPML-II, where the attenuation factor, scale factor, and frequency shift factor formulas are as follows:

$$\alpha = K(\alpha_b + \alpha_e), \alpha_b = (l/L)^{n_\alpha}, \alpha_e = \gamma \exp(-\delta L/l)$$
$$K = \ln(1/R)\frac{(n+1)\sqrt{\mu/\rho}}{2L} \qquad (39)$$
$$\beta = 1 + (\beta_0 - 1)(l/L)^{n_\beta}, \eta = \eta_0 \pi f \left[1 - (l/L)^{n_\eta}\right]$$

Where, L is the RPML thickness, R is the theoretical refection coefcient, l is the distance between the target point and inner boundary. The attenuation factor is an exponential gradient controllable decay function proposed by Luo(Luo and Liu, 2018). The stability factor P is a finger less than 1, the larger the value, the better the absorption effect, the higher the stability, but it hinders the entry of waves into the boundary, causing stronger false reflections.

Next, we re-calculate the RPML-II residual equation using equation 38, but the introduction of scale factor will make the residual equation more complex, with two partial derivatives for time. The RPML-II residual equation is difficult to be generalized to high-order schemes in time like NPML Eq. 35. For this reason, we need to perform special processing on scale factor, rewrite formula 7, and convert it to the time domain.

$$\partial_t \rho v_x = \partial_x \left[\beta_x^{-1}(\tau_{xx} - \varepsilon_{xx}^x)\right] + \partial_z \left[\beta_z^{-1}(\tau_{zx} - \varepsilon_{zx}^z)\right]$$
$$\partial_t \rho v_z = \partial_x \left[\beta_x^{-1}(\tau_{zx} - \varepsilon_{zx}^x)\right] + \partial_z \left[\beta_z^{-1}(\tau_{zz} - \varepsilon_{zz}^z)\right]$$
$$\partial_t \tau_{xx} = C_{11}\partial_x \left[\beta_x^{-1}(v_x - \varepsilon_x^x)\right] + C_{13}\partial_z \left[\beta_z^{-1}(v_z - \varepsilon_z^z)\right] + C_{15}\left(\partial_z \left[\beta_z^{-1}(v_x - \varepsilon_x^z)\right] + \partial_x \left[\beta_x^{-1}(v_z - \varepsilon_z^x)\right]\right) \qquad (40)$$
$$\partial_t \tau_{zz} = C_{13}\partial_x \left[\beta_x^{-1}(v_x - \varepsilon_x^x)\right] + C_{33}\partial_z \left[\beta_z^{-1}(v_z - \varepsilon_z^z)\right] + C_{35}\left(\partial_z \left[\beta_z^{-1}(v_x - \varepsilon_x^z)\right] + \partial_x \left[\beta_x^{-1}(v_z - \varepsilon_z^x)\right]\right)$$
$$\partial_t \tau_{zx} = C_{15}\partial_x \left[\beta_x^{-1}(v_x - \varepsilon_x^x)\right] + C_{35}\partial_z \left[\beta_z^{-1}(v_z - \varepsilon_z^z)\right] + C_{55}\left(\partial_z \left[\beta_z^{-1}(v_x - \varepsilon_x^z)\right] + \partial_x \left[\beta_x^{-1}(v_z - \varepsilon_z^x)\right]\right)$$

Eqs. 40 is the governing equations of the rewritten MCFS-RPML-II and we take the

backward induction method based on NPML to obtain the residual equation, and the process is as follows:

$$\begin{aligned} S_m^{-1}\tilde{\tau}_{nm} &= \beta_m^{-1}\left(\tilde{\tau}_{nm} - \tilde{\varepsilon}_{nm}^m\right) \\ S_m^{-1}\tilde{v}_n &= \beta_m^{-1}\left(\tilde{v}_n - \tilde{\varepsilon}_n^m\right) \end{aligned} \tag{41}$$

Substituting Eqs. 38 and transforming to time domain one obtains:

$$\begin{aligned} \| PML : \partial_t \tilde{\varepsilon}_n^m + P^{n/m}\alpha_m^m \tilde{\varepsilon}_n^m &= \alpha_m^m \tilde{v}_n \\ \partial_t \tilde{\varepsilon}_{nm}^m + P^{n/m}\alpha_m^m \tilde{\varepsilon}_{nm}^m &= \alpha_m^m \tilde{\tau}_{nm} \\ \perp PML : \partial_t \tilde{\varepsilon}_n^m + \left(\eta_m + \beta_m^{-1}\alpha_m^m\right)\tilde{\varepsilon}_n^m &= \beta_m^{-1}\alpha_m^m \tilde{v}_n \\ \partial_t \tilde{\varepsilon}_{nm}^m + \left(\eta_m + \beta_m^{-1}\alpha_m^m\right)\tilde{\varepsilon}_{nm}^m &= \beta_m^{-1}\alpha_m^m \tilde{\tau}_{nm} \end{aligned} \tag{42}$$

According to the above equation, the form of the residual equation of MCFS-RPML-II is consistent with RPML-II, which is equivalent to updating the coefficients by using the frequency shift factor and scale factor according to Eqs. 42 to realize the transition from RPML-II to MCFS-RPML-II.

$$\begin{aligned} C - RPML - \mathrm{II} &: \partial_t \varepsilon^m + A_R \varepsilon^m = B_R v_m \\ C - ADEPML &: \partial_t Q_m^{v_m} + A_A Q_m^{v_m} = -B_A \partial_m v_m \\ C - NPML &: \partial_t \bar{v}_m^n + A_N \bar{v}_m^n = B_N v_m + D_N \partial_t v_m \\ C - RIPML &: \int_0^t A_{RI} M_m^n dt = \int_0^t B_{RI} v_m dt + \partial_n v_m - D_{RI} M_m^n \end{aligned} \tag{43}$$

and

$$\begin{aligned} A_R &= \alpha_m \beta_m^{-1} + \eta_m, B_R = \alpha_m \beta_m^{-1} \\ A_A &= \alpha_m \beta_m^{-1} + \eta_m, B_A = \alpha_m \left(\beta_m^{-1}\right)^2 \\ A_N &= \left(\eta_n \beta_n + \alpha_n\right)\beta_n^{-1}, B_N = \eta_n \beta_n^{-1}, D_N = \beta_n^{-1} \\ A_{RI} &= \beta_n \eta_n + \alpha_n, B_{RI} = \eta_n \alpha_n, D_{RI} = \beta_n \end{aligned} \tag{44}$$

Eqs. 43 summarizes the auxiliary equation forms of four PMLs. Compared with other MCFS-PMLs, the auxiliary equation form of MCFS-RPML-II is the simplest. The other three PMLs all have two partial derivatives or integral terms. After discretization, the RPML-II can be regarded as a simple superposition of variables and residuals at a certain time to obtain the residuals at the next time, which is simple to implement, efficient to calculate, and easy to generalize.

RESULTS AND DISCUSSION

RPML-II has the advantages of simplicity and easy generalization. To validate and investigate its absorption capabilities, computational proficiency, and stability, we evaluated its performance in conjunction with ADEPML and NPML. With no additional elements introduced based on CCS transformation, the performance discrepancies of the three boundaries depend solely on the calculation approach of transforming the governing equation for frequency domain to time domain. Therefore, the performance of the three PMLs can be intuitively judged. The three PMLs are extremely unstable in TTI medium presented in this paper. The stability can be judged by the occurrence time of the instability phenomenon and the enhancement trend of the energy curve. However, their absorption effects need to be studied in a stable environment, so we also provide the energy decay curves in isotropic media.

**Absorption effect and stability study (ADEPML,NPML,RPML-II)**

Complex underground models require higher stability of PMLs. Taking the Marmousi model and TTI medium as examples, we observed the stability of three PMLs.

ADEPML and NPML are currently widely used PMLs, which may exhibit instability when simulating the Marmousi model in elastic isotropic media for a long time. Figure 3 clearly shows that the energy divergence accumulates at the left boundary and ultimately propagates back to the physical region, contaminating the simulated wave field.

The energy decay curve also demonstrated the instability of all three PMLs. System instability occurred at the turning point when energy accumulation in the boundary transmitted back to the physical region, resulting in polluted wave fields. The cut-off time of wave simulation corresponded to the turning point in the physical region.Judging the stability of the three PMLs according to the cut-off time, it is obvious that NPML is less stable, ADEPML is second, and RPML-II has the best stability because the instability phenomenon occurs the latest, and the energy is greatly suppressed compared with the other two PMLs.

In order to further study the stability and absorption performance, a large angle incident model is established. The time step used is 0.5 ms, the grid spacing is 5 m, the model size is 2500×1000 m.10 absorption layers are added, the source is applied at (1300,55)m, and the source-time function is a Ricker wavelet with a center frequency of 25 Hz. Because the TTI medium given will produce shear wave crossing phenomenon, the complex shear wave field is difficult to be absorbed after entering the boundary layer, which will soon cause instability. Instability occurs too quickly, making it difficult to effectively observe the absorption effect of the three boundaries. Therefore, based on this model, we modify the elastic parameters and carry out wave field simulation in elastic isotropic media. The parameters of the two models are shown in Table 1.

As shown in Figures 4a-f, the three PMLs are difficult to absorb the splitting shear waves, resulting in energy accumulation, which is rapidly transmitted back to the physical region. In Figure 4h, its energy decay curve also shows that instability phenomenon occurs rapidly, even without energy decay section,but directly increases exponentially. The results in Figure 4h are consistent with those under the Marmousi model, and RPML-II has better stability.

Figure 4g shows the energy decay curve in isotropic medium. The absorption effects of NPML and RPML-II are similar, with a consistent trend observed in the energy decay curve. The attenuation curves of NPML and RPML-II, represented by the arrow, are lower than that of ADEPML, indicating their superior absorption performance. However, NPML and RPML-II exhibit a higher amount of residual energy in the absorbing layer, which leads to the lower stability of NPML as compared to ADEPML. Excessive residual energy will inevitably affect the stability of RPML-II, and long-term numerical simulation experiments have also shown that even if there is a large amount of energy in the boundary, RPML-II still has better stability.

**Improvement of RPML-II (MCFS-RPML-II)**

In the previous section, we studied the absorption effect and stability of ADEPML, NPML, and RPML-II, indicating that the loading and discretization methods of RPML are more stable and cause weaker false reflections. RPML-II is still difficult to absorb grazing waves. Although RPML-II is more stable than the other two PMLs, it still cannot be practically applied to TTI media or some complex model wave field simulations. In order to address the limitations of conventional PML, we have proposed MCFS-RPML- II.

For MCFS-RPML-II, stability factor, frequency shift factor, and scale factor are further introduced on the basis of attenuation factor. We investigate the effects of various factors on RPML-II through numerical simulation.

In Figure 5, MCFS-RPML-II is simplified to RPML-II when the stability factor is set to 0, the scale factor is set to 1, and the frequency shift factor is set to 0. Parameters that are not individually marked in the figure are evaluated as described above. As shown in wavefield snapshot, the introduction of all three factors favors the absorption of energy inside the boundary. The larger the stability factor value, the better the boundary absorption effect.

Because the stability factor determines the strength of the parallel attenuation profile, the strong attenuation profile can better absorb the dissipation wave propagating along the boundary direction. However, the introduction of parallel profiles will hinder the entry of waves into the boundary, causing stronger false reflections, so the stability factor value should not be too large.

The scale factor can bend the incident wave towards normal direction, allowing the wave to propagate deeper into the boundary, and plays an important role in absorption of high angle incident waves. The addition of the frequency shift factor serves to prevent the appearance of low-frequency singular values. From the attenuation curve, it can be seen that the scale factor delays the generation of instability, and the frequency shift factor makes the system relatively stable, without significant instability observed. Nevertheless, there are still local energy oscillations in attenuation curve(Figure 6d). Two factors can further enhance the stability while weakening the internal energy of the boundary. CFSPML still faces stability issues, and the introduction of stability factors makes the system asymptotically stable. When the stability factor is set to 0.1, curve shows that the system is stable and there is no instability phenomenon during long-term simulation. However, this value can cause strong false reflections, as shown in Figure 5h. Therefore, while ensuring stability, try to select a smaller stability factor value as much as possible.

For MCFS-RPML-II in Figure 5b and Figure 6a, we selected the $\beta_0$ as 3, $\eta_0$ as 2, and P

as 0.02. These parameters were chosen to improve absorption performance and stability, while preventing false reflections caused by excessively high stability factor values and minimizing local instability in CFS-RPML-II. RPML-II and NPML retain a large amount of energy in the boundary, which is not suppressed by simply introducing frequency shift and scale factors. The larger the stability factor value, the smaller the difference between the solid line and dashed curves, and the energy is further absorbed. Therefore, the introduction of stability factor is essential for RPML-II. As shown in Figure 6a and Figure 6c, when three factors are introduced simultaneously, even if the stability factor is 0.02, the difference between the dashed and solid lines is similar to the value of 0.1.

The model parameters in Figure 7 are shown in Table I. Low-velocity anomalies are developed in anticline structures, and their wave field snapshots and energy attenuation curves clearly demonstrate the good absorption effect of MCFS-RPML-II.

**Research on Computational efficiency (ADEPML,NPML,RPML-II)**

RPML-II has the most concise auxiliary equation format, requiring only one time difference, and theoretically has the highest computational performance. To this end, under the same conditions, we obtained the forward iterative computation time of the above three model cases. The global PML method is employed, wherein the attenuation factor in the physical region is set to 0 for simplicity of implementation.

We present the system time under three PMLs and conducted statistical analysis on isotropic, anisotropic, and anticline models, respectively. The results correspond to Figure 4g, Figure 5, and Figure 7. For precise calculation time, we have recorded the total time taken for

30,000 iterations. As portrayed in Figure 8, computational time consumption for the three PMLs is consistent across all three models, with $t_{ADEPML} > t_{NPML} > t_{RPML}$. The auxiliary equation of ADEPML has a spatial derivative term, and this article adopts a spatial twelfth order discretization method. This makes the cost of spatial derivative discretization higher than that of temporal derivative term. NPML includes two time derivative terms, while RPML-II only contains one, making it the most efficient.

## CONCLUSIONS

The process of obtaining RPML-II residuals involves establishing connections between multiple PMLs. Upon comparing and analyzing the performance of various PMLs, it was found that RPML-II has the most concise auxiliary equation, which retains the original equation form and leads to better computational performance and wider application. The MCFS-RPML-II proposed by special processing does not complicate the auxiliary equation due to the introduction of three factors.The form of the auxiliary equation remains unchanged, and only the coefficients can be modified. In numerical simulation, we found that RPML-II has higher stability than NPML and ADEPML, and its absorption performance is similar to NPML, superior to ADEPML, and its computational performance is also the highest.

RPML-II, like NPML, does not change the original equation form, which is conducive to generalization, and their absorption performance is similar. However, RPML has stronger stability and higher computational efficiency, and is easier to generalize to temporal high-order wave field simulation.Compared to ADEPML, RPML-II has higher absorption performance, stability, and computational efficiency, but its ability to suppress energy in the boundary is insufficient, resulting in excessive residual energy. This issue was resolved in the subsequent MCFS-RPML-II derivation. MCFS-RPML-II further introduces stability factor, frequency shift factor and scale factor on the basis of RPML-II to improve boundary stability and absorption effect. Numerical simulation experiments indicate that the scale factor and frequency shift factor enhance the absorption of high-angle incident waves, contributing to stability. The stability factor further absorbs residual energy, suppressing local instability phenomena in CFS-RPML-II. Conventional PMLs, including RPML-II with high stability, are difficult to meet the demands of anisotropic wave field simulation, while MCFS-RPML-II exhibits higher absorption performance, stability, and computational efficiency in complex media and model wave field simulation.

REFERENCES

Antoine, X., and X. F. Zhao, 2022, Pseudospectral methods with PML for nonlinear Klein-Gordon equations in classical and non-relativistic regimes: Journal of Computational Physics, 448, 110728.

Appelö, D., and G. Kreiss, 2006, A new absorbing layer for elastic waves: Journal of Computational Physics, 215, 642–660.

Appelö, D., and T. Colonius, 2009, A high-order super-grid-scale absorbing layer and its application to linear hyperbolic systems: Journal of Computational Physics, 228, no. 11, 4200–4217.

Bécache, E., S. Fauqueux, and P. Joly, 2003, Stability of perfectly matched layers, group velocities and anisotropic waves: Journal of Computational Physics, 188, no. 2, 399–433.

Bécache, E., P. Joly, and M. Kachanovska, 2017a, Stable perfectly matched layers for a cold plasma in a strong background magnetic field: Journal of Computational Physics, 341, 76–101.

Bécache, E., P. Joly, and V. Vinoles, 2017b, On the analysis of perfectly matched layers for a class of dispersive media and application to negative index metamaterials: Mathematics of Computation, 87, 2775–2810.

Berenger, J. P., 1994, A perfectly matched layer for the absorption of electromagnetic waves: Journal of Computational Physics, 114, 185-200.

Berenger, J. P., 1996, Perfectly matched layer for the FDTD solution of wave-structure interaction problems: IEEE Transactions on Antennas and Propagation, 44, 110–117.

Cerjan, C., D. Kosloff, R. Kosloff, et al., 1985, A nonreflecting boundary condition for discrete acoustic and elastic wave equations: Geophysics, 50, 705-708.

Chen, J. Y., 2012, Nearly perfectly matched layer method for seismic wave propagation in poroelastic media: Canadian Journal of Exploration Geophysics, 37, 22-27.

Chew, W. C., and Q. H. Liu, 1996, Perfectly matched layers for elastodynamics: A new absorbing boundary condition: Journal of Computational Acoustics, 4, 341-359.

Clayton, R., and B. Engquist, 1977, Absorbing boundary conditions for acoustic and elastic wave equations: Bulletin of the Seismological Society of America, 67, 1529-1540.

Collino, F., and C. Tsogka, 2001, Application of the PML absorbing layer model to the linear elastodynamic problem in anisotropic heterogeneous media: Geophysics, 66, 294-307.

Cummer, S. A., 2003, A simple nearly perfectly matched layer for general electromagnetic media: IEEE Microwave and Wireless Components Letters, 13, no. 3, 128-130.

Cummer, S., 2004, Perfectly matched layer behavior in negative refractive index materials: IEEE Antennas and Wireless Propagation Letters, 3, 172–175.

Demaldent, E., and S. Imperiale, 2013, Perfectly matched transmission problem with absorbing layers: Application to anisotropic acoustics in convex polygonal domains: International Journal for Numerical Methods in Engineering, 96, no. 11, 689–711.

Drossaert, F. H., and A. Giannopoulos, 2007, A nonsplit complex frequency-shifted PML based on recursive integration for FDTD modeling of elastic waves: Geophysics, 72, no. 2, T9–T17

Duru, K., 2014, A perfectly matched layer for the time-dependent wave equation in heterogeneous and layered media: Journal of Computational Physics, 257, 757–781

Duru, K., J. E. Kozdon, and G. Kreiss, 2015, Boundary conditions and stability of a perfectly matched layer for the elastic wave equation in fifirst order form: Journal of Computational Physics, 303, 372–395.

Engquist, B. and A. Majda, 1977, Absorbing boundary conditions for the numerical simulation of waves: Mathematical Computing, 31, 629-651.

Demaldent, E., and S. Imperiale, 2013, Perfectly matched transmission problem with absorbing layers: Application to anisotropic acoustics in convex polygonal domains: International Journal for Numerical Methods in Engineering, 96, no. 11, 689–711.

Dong, S. Q., L. G. Han, Y, Hu, and Y. Yin, 2020, Full waveform inversion based on a local traveltime correction and zero-mean cross-correlation-based misft function: Acta Geophysica, 2020, 68 no. 1, 29-50.

Feng, N., Y. Yue, C. Zhu, L. Wan, and Q. H. Liu, 2015, Second-order pml: Optimal

choice of nth-order pml for truncating fdtd domains: Journal of Computational Physics, 285, 71-83.

Gedney, S. D., and B. Zhao, 2010, An Auxiliary Differential Equation Formulation for the ComplexFrequency Shifted PML: IEEE Transactions on Antennas and Propagation, 58, no. 3, 838-847.

Giannopoulos, A., 2018, Multipole Perfectly Matched Layer for Finite-Difference Time-Domain Electromagnetic Modeling: IEEE Transactions on Antennas and Propagation, 66, no. 6, 2987–2995.

Guo, G. S., S. Operto, A. Gholami, and H. S. Aghamiry, 2024, Time-domain extended-source full-waveform inversion: Algorithm and practical workflow: Geophysics, 89, no. 2, R73–R94, doi:10.1190/geo2023-0055.1.

Higdon, R. L., 1991, Absorbing boundary conditions for elastic waves: Geophysics, 56, 231-241.

Harari, I., M. Slavutin, and E. Turkel, 2000, Analytical and numerical studies of a finite element PML for the helmholtz equation: Journal of Computational Acoustics, 8, 121-137.

Hastings, F. D., J. B. Schneider, and S. L. Broschat, 1996, Application of the perfectly matched layer(PML) absorbing boundary condition to elastic wave propagation: Journal of the Acoustical Society of America, 100, 3061-3069.

Hu, F. Q., 1996, On Absorbing Boundary Conditions for Linearized Euler Equations by a Perfectly Matched Layer: Journal of Computational Physics, 129, 201-219.

Hu, F. Q., 2001, A Stable, Perfectly Matched Layer for Linearized Euler Equations in Unsplit Physical Variables: Journal of Computational Physics, 173(2), 455–480.

Huang, J. D., D. H. Yang, X. J. He, J.K Sui, and S. L. Liang, 2023, Double-pole unsplit complex-frequency-shifted multiaxial perfectly matched layer combined with strong-stability-preserved Runge-Kutta time discretization for seismic wave equation based on the discontinuous Galerkin method: Geophysics, 88, no. 5, T259–T270, doi:10.1190/geo2022-0776.1.

Katz, D. S., E. T. Thiele, and A. Taflflove, 1994, Validation and extension to three dimensions of the Berenger PML absorbing boundary condition for FD-TD meshes: IEEE Microwave and Guided Wave Letters, 4, 268–270.

Komatitsch, D., and R. Martin, 2007, An unsplit convolutional perfectly matched layer improved at grazing incidence for the seismic wave equation: Geophysics, 72, no. 5, 155-167.

Kuzuoglu, M., and R. Mittra, 1996, Frequency dependence of the constitutive parameters of causal perfectly matched anisotropic absorbers: IEEE Microwave and Guided Wave Letters, 6, no. 12, 447-449.

Lei, D., L. Y. Yang, C. M. Fu, R. Wang, and Z. X. Wang, 2022, The application of a novel perfectly matched layer in magnetotelluric simulations: Geophysics, 87, no. 3, E163–E175, doi:10.1190/geo2020-0393.1.

Li, J., K. A. Innanen, and B. Wang, 2019, A New Second Order Absorbing Boundary Layer Formulation for Anisotropic-Elastic Wavefield Simulation: Pure and Applied Geophysics, 176, no. 4, doi:10.1007/s00024-018-2046-z.

Luo, Y. Q., and C. Liu, 2018, Absorption effects in nearly perfectly matched layers and damping factor improvement: OGP, 53, no. 5, 903-913.

Luo, Y. Q., and C. Liu, 2020, On the stability and absorption effect of the multiaxial complex frequency shifted nearly perfectly matched layers method for seismic wave propagation: Chinese Journal of Geophysics (in Chinese), 63, no. 8, 3078-3090.

Luo, Y. Q., and C. Liu,, 2022, Modeling seismic wave propagation in TTI media using multiaxial complex frequency shifted nearly perfectly matched layer method: Acta Geophysica, 70, no. 1, 89-109.

Ma, X., D. H. Yang, X. Y. Huang, and Y. J. Zhou, 2018, Nonsplit complex-frequency shifted perfectly matched layer combined with symplectic methods for solving second-order seismic wave equations — Part 1: Method: Geophysics, 83, no. 6, T301–T311.

Ma, X., D. H. Yang, X. Y. Huang, and Y. J. Zhou, 2019, Nonsplit complex-frequency-shifted perfectly matched layer combined with symplectic methods for solving second-order seismic wave equations — Part 2: Wavefield simulations: Geophysics, 84, no. 3, T167–T179, doi:10.1190/GEO2018-0349.1.

Martin, R., D. Komatitsch, and S. D. Gedney, 2010, et al. A high-order time and space formulation of the unsplit perfectly matched layer for the seismic wave equation using Auxiliary Differential Equations (ADE-PML): Computer Modeling in Engineering & Sciences, 56, no. 1, 17-41.

Mezafajardo, K. C., and A. S. Papageorgiou, 2008, A Nonconvolutional, Split-Field, Perfectly Matched Layer for Wave Propagation in Isotropic and Anisotropic Elastic Media: Stability Analysis: Bulletin of the Seismological Society of America, 98, no. 4, 1811-1836.

MezaFajardo, K. C., and A. S. Papageorgiou, 2010, On the stability of a non-convolutional perfectly matched layer for isotropic elastic media: Soil Dynamics & Earthquake Engineering, 30, no. 3, 68-81.

Nataf, F., 2005, A new construction of perfectly matched layers for the linearized Euler equations: Comptes Rendus Mathematique, 340, 775-778.

Nissen, A., and G. Kreiss, 2011, An Optimized Perfectly Matched Layer for the Schrödinger Equation: Communications in Computational Physics, 9, 147-179.

Ramadan, O., and A. Y. Oztoprak, 2002, DSP techniques for implementation of perfectly matched layer for truncating FDTD domains: Electronics Letters, 38, no. 5, 211-212.

Reynolds, A. C.,1978, Boundary conditions for the numerical solution of wave propagation problems: Geophysics, 43, 895-904.

Roden, J. A., and S. D. Gedney, 2000, Convolution PML (C-PML):An efficient FDTD implementation of the CFS-PML for arbitrary media: Microwave and Optical Technology Letters, 27, no. 5, 334-339.

Sochacki, J., R. Kubichek, and J. George, 1987, Absorbing boundary conditions and surface waves: Geophysics, 52, 60-71.

Teixeira, F. L., and W. C. Chew, 1998, Extension of the PML absorbing boundary condition to 3D spherical coordinates: scalar case: IEEE Transactions on Magnetics, 34, 2680-2683.

Wang, E. J., J. M. Carcione, J. Ba, M. Alajmi, and A. N. Qadrouh, 2019, Nearly perfectly matched layer absorber for viscoelastic wave equations: Geophysics, 84, no. 5, T335–T345, doi:10.1190/geo2018-0732.1.

Wang, J., Y. Wang, and D. Zhang, 2006, Truncation of open boundaries of cylindrical waveguides in 2.5-dimensional problems by using the convolutional perfectly matched layer: IEEE Transactions on Plasma Science, 34, 681-690.

Wen, L., Y. Liu, G. Li, S. Zhu, G. X. Chen, and C. F. Li , 2023, 2D frequency-domain finite-difference acoustic wave modeling using optimized perfectly matched layers, Geophysics, 88, no. 2, F1–F13, doi:10.1190/geo2022-0145.1

Wu, H., X. T. Dong,T. Zhong, S. K. Zhang, and S. P. Lu, 2023, Least-squares Reverse Time Migration using the Inverse Scattering Imaging Condition: IEEE Transactions on Geoscience and Remote Sensing, 61, 1-12.

Wu, Y. Q., H. S. Aghamiry, S. Operto, and J. W. Ma, 2023, Helmholtz-equation solution in nonsmooth media by a physics-informed neural network incorporating quadratic terms and a perfectly matching layer condition: Geophysics, 88, no. 4, T185–T202,doi:10.1190/geo2022-0479.1.

Zeng, Y. Q., J. Q. He, and Q. H. Liu, 2001, The application of the perfectly matched layer in numerical modeling of wave propagation in poroelastic media: Geophysics, 66, 1258-1266.

Zhang, T. T., and X. K. Li, 2023, Stability of Perfectly Matched Layers for Time Fractional Schrödinger Equation: Engineering, 15, 1-12.

LIST OF FIGURES

Fig.1 Diagram of relationships between different PMLs.

Fig.2 Schematic of RPML and MCFS-RPML.

Fig.3 Wave field snapshots and energy decay curves under the marmousi model.(a)Marmousi model. (b-d)Wave Field Snapshots using ADEPML, RPML and NPML, respectively. (e)The energy decay curves under three distinct boundary conditions.

Fig.4. Wave field snapshot and energy attenuation curve. (a,c and e) 0.2s TTI medium. (b,d and f) 0.6s TTI medium. (g) Isotropic medium. (h) TTI medium. (a and b) Wave Field Snapshots using ADEPML. (a and b) Wave Field Snapshots using NPML.(a and b) Wave Field Snapshots using RPML.

Fig.5. Snapshots of the wave field under different parameters. (a)RPML,$\beta_0=1,\eta_0=0$, P=0. (b)MCFS-RPML,$\beta_0=3,\eta_0=2$, P=0.02. (c)CFS-RPML,$\beta_0=3,\eta_0=0$, P=0. (d)CFS-RPML, $\beta_0=5,\eta_0=0$, P=0. (e)CFS-RPML,$\beta_0=1,\eta_0=2$, P=0. (f)CFS-RPML,$\beta_0=1,\eta_0=3$, P=0. (g)M-RPML,$\beta_0=1,\eta_0=0$, P=0.02. (h)M-RPML,$\beta_0=1,\eta_0=0$, P=0.1.

Fig.6. Energy attenuation curves under different parameters.(a) RPML,MCFS-RPML. (b)CFS-RPML. (c)M-RPML. (d)CFS-RPML.

Fig.7. Wave Field Snapshots and Energy Decay Curves under the Anticline Model.(a)Anticline Model. (b-c)Wave Field Snapshots.(d) Energy Decay Curves.

Fig.8. Calculation time under different models.

Table 1. Model parameters

|  | ρ(Kg/m³) | C11(10⁹N/m²) | C33 | C44 | C66 | C13 |
|---|---|---|---|---|---|---|
| TTI medium | 2300 | 26.40 | 15.60 | 4.38 | 6.84 | 6.11 |
| Anticline model 1 | 2300 | 26.40 | 15.60 | 4.38 | 6.84 | 6.11 |
| Anticline model 2 | 2420 | 36.53 | 21.49 | 5.63 | 8.78 | 9.24 |
| Anticline model 3 | 2100 | 15.68 | 4.60 | 2.45 | 3.67 | 4.60 |

|  | ρ(Kg/m³) | Vp(m/s) | Vs(m/s) |
|---|---|---|---|
| Isotropic medium | 2000 | 3000 | 1400 |

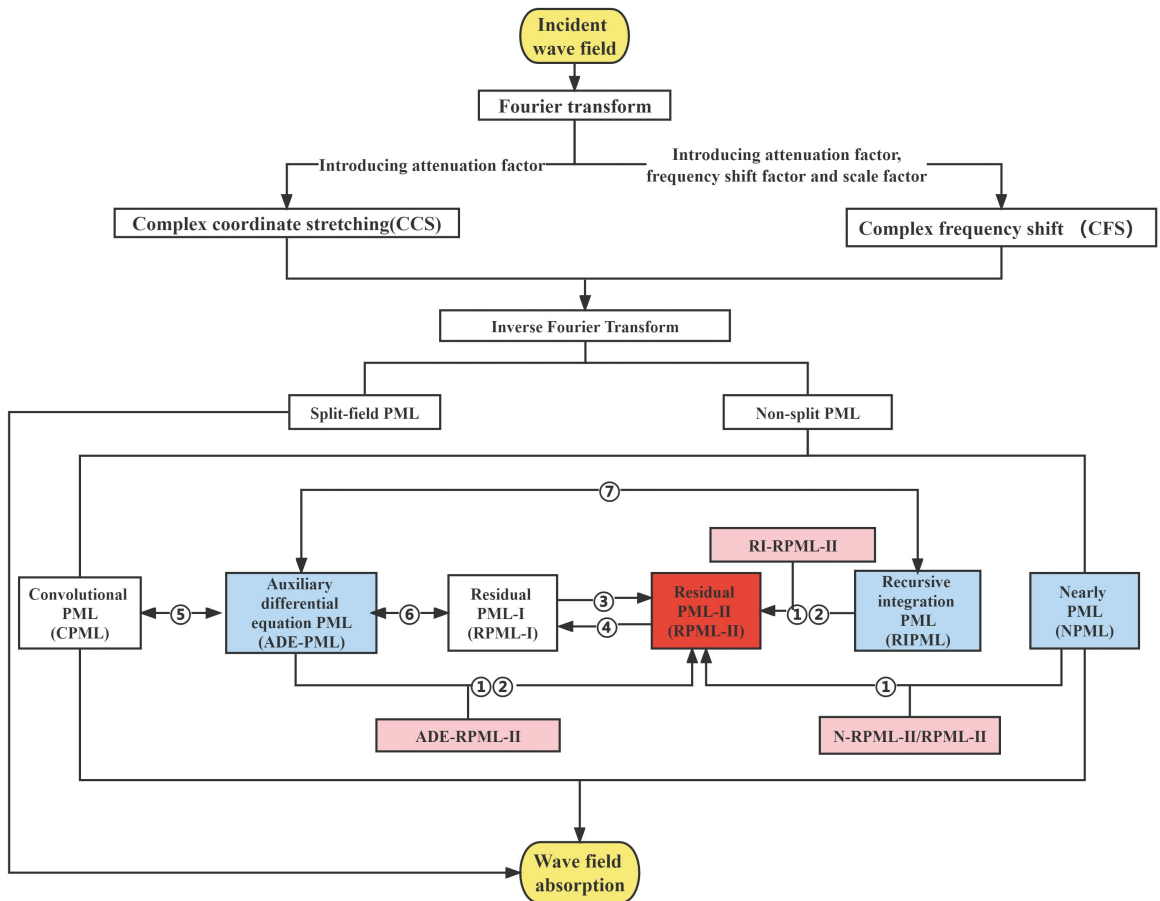

Figure 1. Diagram of relationships between different PMLs.

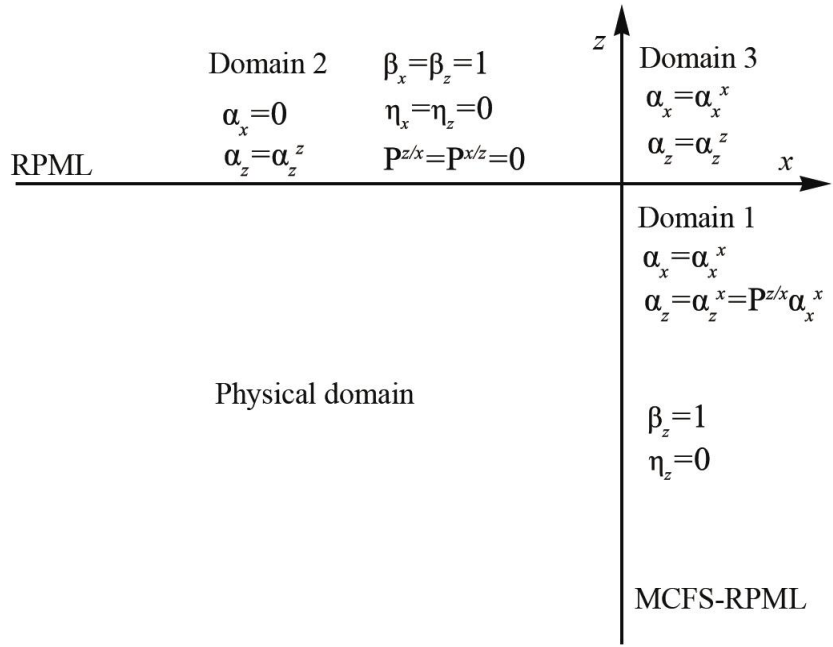

Figure 2. Schematic of RPML and MCFS-RPML.

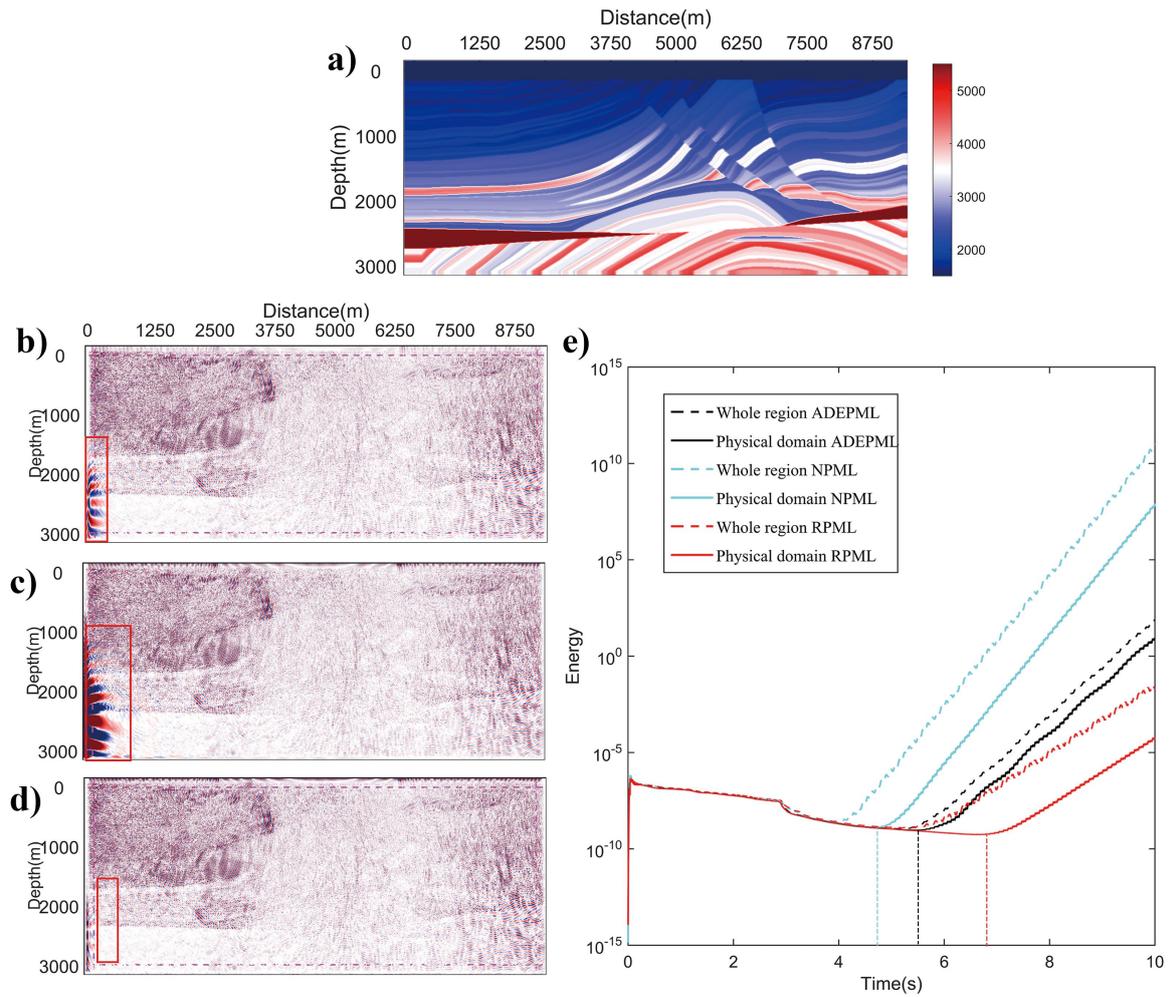

Figure 3 Wave field snapshots and energy decay curves under the marmousi model.(a)Marmousi model. (b-d)Wave Field Snapshots using ADEPML, RPML and NPML, respectively. (e)The energy decay curves under three distinct boundary conditions.

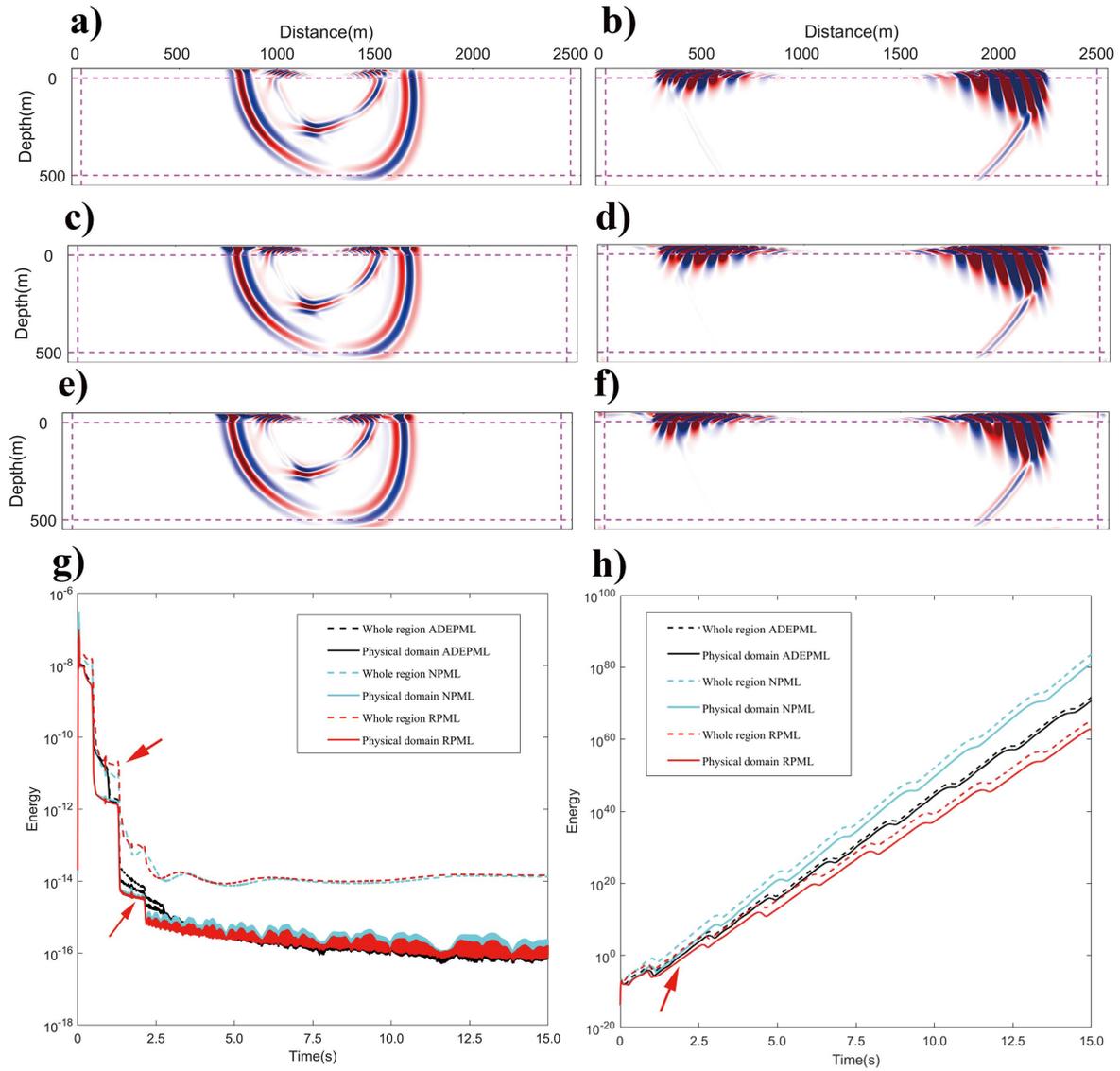

Figure 4. Wave field snapshot and energy attenuation curve. (a,c and e) 0.2s TTI medium. (b,d and f) 0.6s TTI medium. (g) Isotropic medium. (h) TTI medium. (a and b) Wave Field Snapshots using ADEPML. (a and b) Wave Field Snapshots using NPML.(a and b) Wave Field Snapshots using RPML.

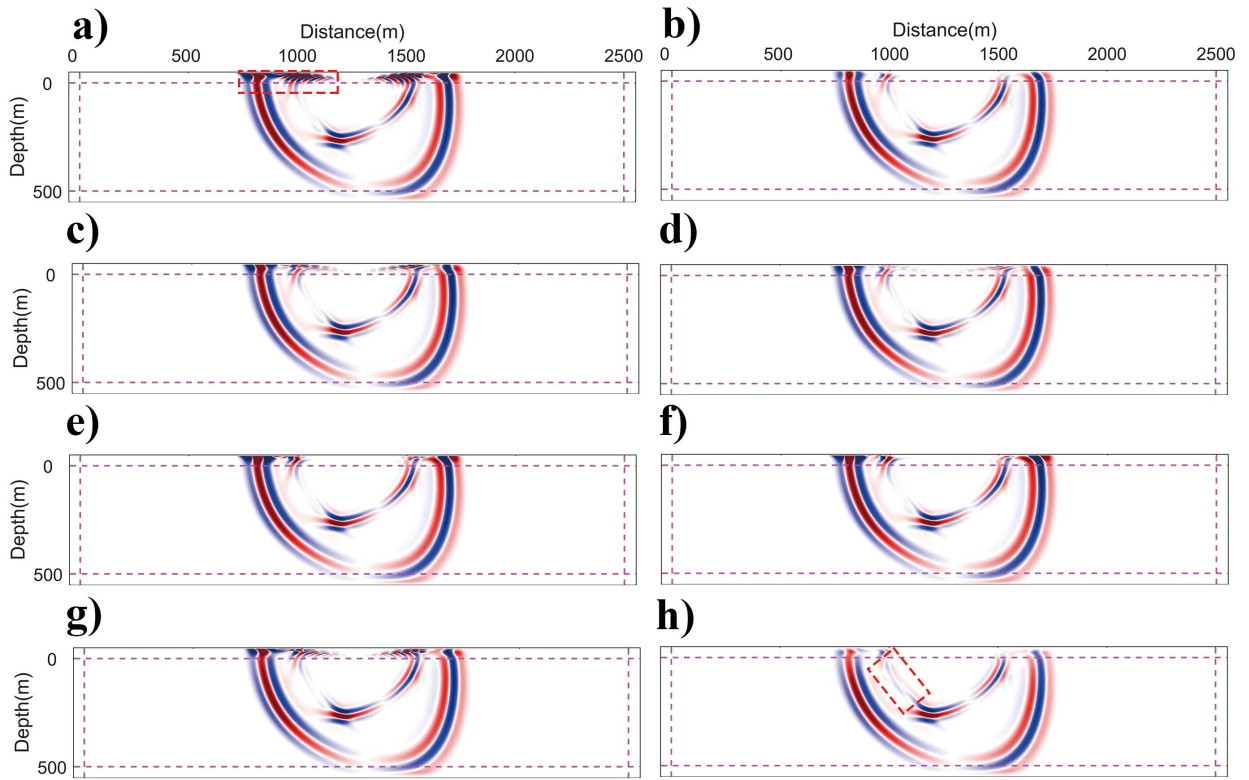

Figure 5. Snapshots of the wave field under different parameters. (a)RPML,$\beta_0=1$,$\eta_0=0$, P=0. (b)MCFS-RPML,$\beta_0=3$,$\eta_0=2$, P=0.02. (c)CFS-RPML,$\beta_0=3$,$\eta_0=0$, P=0. (d)CFS-RPML, $\beta_0=5$,$\eta_0=0$, P=0. (e)CFS-RPML,$\beta_0=1$,$\eta_0=2$, P=0. (f)CFS-RPML,$\beta_0=1$,$\eta_0=3$, P=0. (g)M-RPML,$\beta_0=1$, $\eta_0=0$, P=0.02. (h)M-RPML,$\beta_0=1$,$\eta_0=0$, P=0.1.

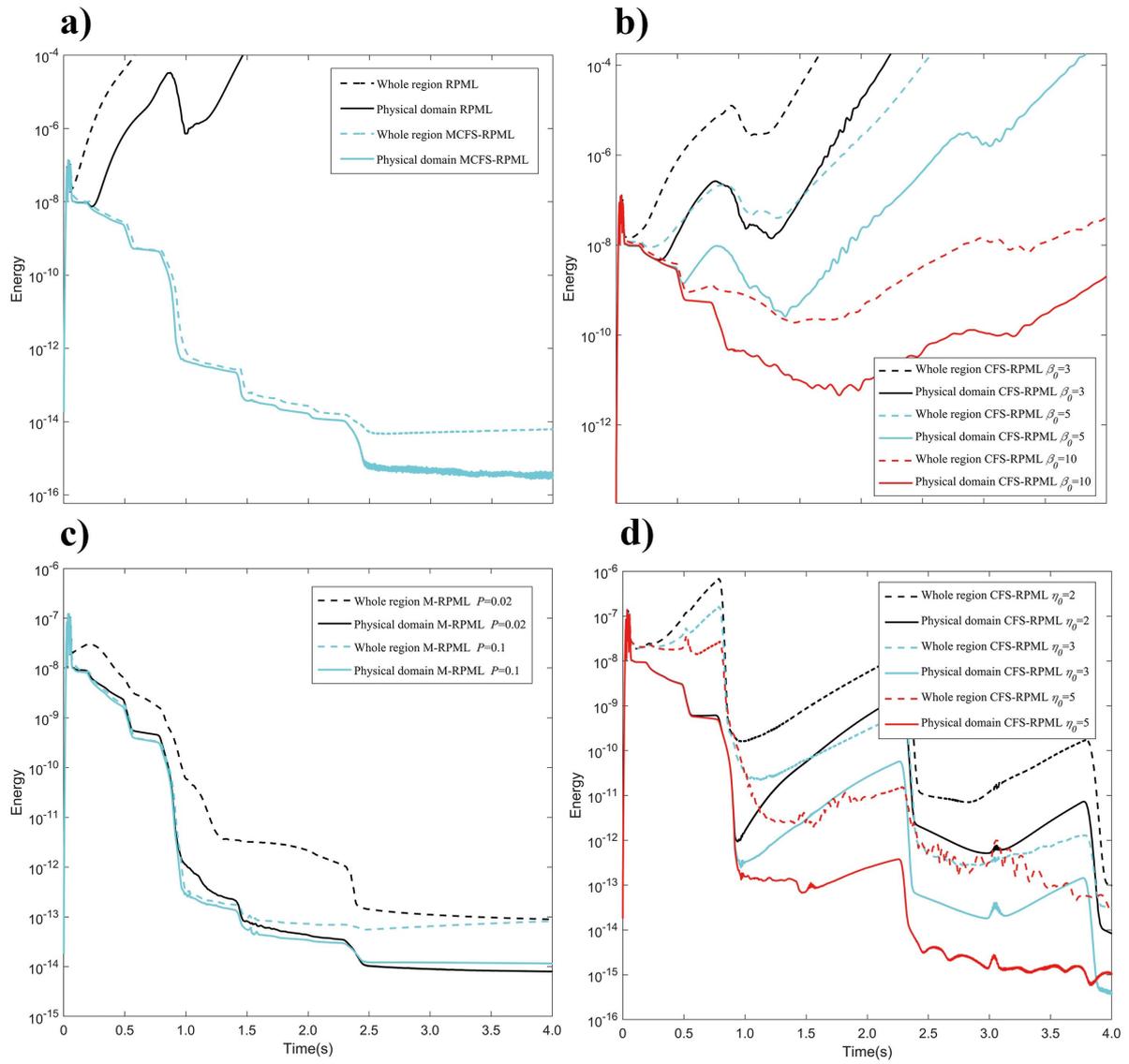

Figure 6. Energy attenuation curves under different parameters.(a) RPML,MCFS-RPML. (b)CFS-RPML. (c)M-RPML. (d)CFS-RPML.

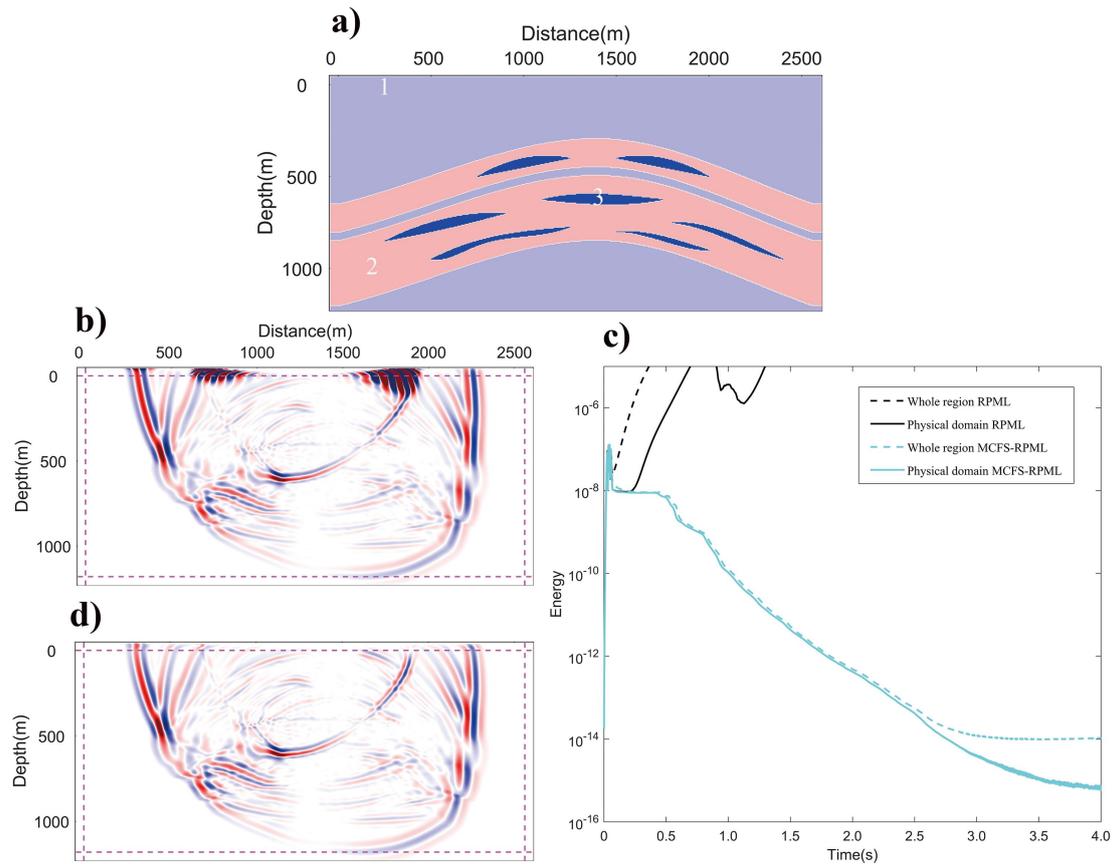

Figure 7. Wave Field Snapshots and Energy Decay Curves under the Anticline Model.(a)Anticline Model. (b-c)Wave Field Snapshots.(d) Energy Decay Curves.

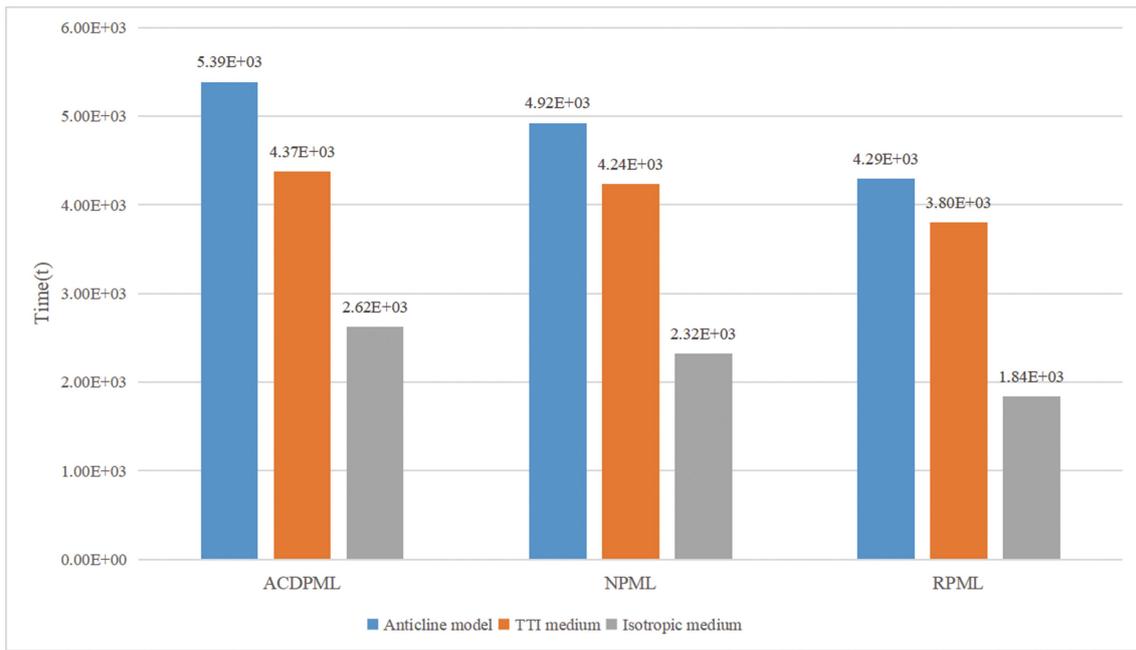

Figure 8. Calculation time under different models.